\documentclass[journal]{IEEEtran}
\ifCLASSINFOpdf
\else
\fi
\usepackage[ruled,vlined]{algorithm2e}
\usepackage{caption}
\usepackage{booktabs}
\usepackage{amsmath,amssymb,amsfonts}
\usepackage{graphicx}
\usepackage{xspace}
\usepackage{textcomp}
\usepackage[nolist]{acronym}
\usepackage{color, colortbl}
\usepackage{adjustbox}
\usepackage[colorinlistoftodos]{todonotes}
\usepackage{xcolor}
\usepackage[bottom, multiple]{footmisc}
\usepackage{pgfplots}
\usepackage{pgfplotstable}
\usepackage{tikz}
\usepackage{subfigure}
\usepackage{multirow}

\usepackage{balance}
\usepackage{comment}
\PassOptionsToPackage{hyphens}{url}\usepackage{hyperref}
\newcommand{\iot}{IoT\xspace}
\newcommand{\lte}{LTE\xspace(CAT-M1/NB-IoT)\xspace}
\newcommand{\system}{\textit{Resilient Edge}\xspace}
\newcommand{\lorawan}{LoRaWAN\xspace}
\newcommand{\lora}{LoRa\xspace}
\newcommand{\sigfox}{Sigfox\xspace}
\newcommand{\nbiot}{NB-IoT\xspace}
\newcommand{\wifi}{Wi-Fi\xspace}
\newcommand{\zigbee}{ZigBee\xspace}

\newcommand{\bluetooth}{Bluetooth\xspace}
\newcommand{\appX}{HealthApp\xspace}
\newcommand{\appY}{HomeApp\xspace}
\newcommand{\fipy}{FiPy\xspace}
\newcommand{\rpi}{RPi\xspace}
\newcommand{\etal}{et al.\@\xspace}

\newcommand{\noindgras}[1]{{\noindent \textbf{#1}}}

\def\BibTeX{{\rm B\kern-.05em{\sc i\kern-.025em b}\kern-.08em
    T\kern-.1667em\lower.7ex\hbox{E}\kern-.125emX}}

\begin{acronym}
\acro{6LoWPAN}{IPv6 over Low Power Wireless Personal Area Networks}
\acro{RF}{Radio frequency}
\acro{RSSI}{Received Signal Strength Indicator}
\acro{LQI}{Link Quality Indicator }
\acro{SCK}{Smart Citizen Kit}
\acro{eCO2}{equivalent CO2}
\acro{UO}{Urban Observatory}
\acro{IoT}{Internet of Things}
\acro{WSN}{Wireless Sensor Networks}
\acro{SHM}{Structural Health Monitoring}
\acro{BSI}{British Standards Institute}
\acro{ICT}{Information and Communications Technology}
\acro{SolarPV}{Solar Photovoltaic Systems}
\acro{KPI}{Key Performance Indicators}
\acro{MCU}{Microcontrollers}
\acro{APNR}{Automatic Plate Number Recognition}
\acro{CCTV}{Closed-circuit television}
\acro{CO}{Carbon Monoxide}
\acro{CO2}{Carbon Dioxide}
\acro{NO}{Nitric Oxide}
\acro{NO2}{Nitrogen Dioxide}
\acro{CH4}{Methane}
\acro{H2S}{Hydrogen sulfide}
\acro{NH3}{Ammonia}
\acro{ECG}{Electrocardiogram}
\acro{LIDAR}{Light Detection and Ranging}
\acro{NOAA}{National Oceanic and Atmospheric Administration}
\acro{LTE}{Long Term Evolution}
\acro{GPRS}{General Packet Radio Service}
\acro{CDMA}{Code-division multiple access}
\acro{GSM}{Global System for Mobile Communication}
\acro{NB-IoT}{Narrowband IoT}
\acro{M2M}{Machine to Machine}
\acro{WiMAX}{Worldwide Interoperability for Microwave Access}
\acro{UWB}{Ultra Wideband}
\acro{BUO}{Bristol Urban Observatory}
\acro{LAN}{Local Area Network}
\acro{RFID}{Radio-frequency identification}
\acro{NFC}{Near-field communication}
\acro{BCG}{Ballistocardiographic}
\acro{PIR}{Passive Infrared Sensor}
\acro{IAQ}{Indoor Air Quality}
\acro{VOC}{Volatile Organic Compounds}
\acro{EV}{Electric Vehicle}
\acro{TfL}{Transport for London}
\acro{TERS}{Temporary earth restraining structure}
\acro{SLA}{Service level agreement}
\acro{LoRaWAN}{Long Range Wide Area Network}
\acro{WiMAX}{Worldwide Interoperability for Microwave Access}
\acro{MINA}{Multinetwork INformation Architecture}
\acro{AoT}{Array of Things}
\acro{OAA}{Observe Analyse Adapt}
\acro{SBC}{Single Board Computer}
\acro{SPHERE}{Sensor Platform for HEalthcare in a Residential Environment}
\acro{SHG}{SPHERE Home Gateway}
\acro{SDH}{SPHERE Data Hub}
\acro{NUC}{Next Unit of Computing}
\acro{LCD}{Liquid Crystal Display}
\acro{SONET}{Synchronous optical networking}
\acro{DSL}{Digital Subscriber Line}
\acro{LED}{Light Emitting Diode}
\acro{SoC}{System on Chip}
\acro{TBI}{Traumatic Brain Injury}
\acro{AURN}{Automatic Urban and Rural Network}
\acro{VOCS}{Volatile Organic Compounds}
\acro{QoL}{Quality of Life}
\acro{AQI}{Air Quality Index}
\acro{UKCRIC}{UK Collaboratorium for Research On Infrastructure and Cities}
\acro{NAK}{Network Authentication Key}
\acro{MAC}{Message Authentication Code}
\acro{APN}{Access Point Names}
\acro{DoNAS}{Data over NAS}
\acro{NIDD}{Non-IP Data Delivery}
\acro{SIM}{Subscriber Identity Modules}
\acro{UNB}{Ultra-Narrow Band}
\acro{ADR}{Adaptive Data Rate}
\acro{QoS}{Quality of Service}
\acro{PSM}{Power Saving Mode}
\acro{ILP}{Integer Linear Programming}
\acro{UE}{User Equipment}
\acro{TTN}{The Things Network}
\acro{eDRX}{Extended Discontinuous reception}
\acro{OTAA}{Over-the-air Activation}
\acro{ABP}{Activation-by-personalization}
\acro{MIC}{Message Integrity Code}
\acro{PLMN}{Public Land Mobile Network}
\acro{DL}{Down link}
\acro{UL}{Up link}
\acro{MTU}{Maximum Transmission Unit}
\acro{TCP}{Transmission Control Protocol}
\acro{IP}{Internet Protocol}
\acro{SF}{Spreading Factor}
\acro{WPA}{Wi-Fi Protected Access}
\acro{WPS}{Wi-Fi Protected Setup}
\acro{SMS}{short message service}
\acro{NTP}{Network Time Protocol}
\acro{DoS}{Denial-of-Service}
\acro{CCTV}{closed-circuit television}
\acro{SIM}{subscriber identity module}
\acro{APN}{Access Point Name}
\acro{REST}{Representational state transfer}
\acro{API}{Application Programming Interface}
\acro{TBS}{Transport block sizes}
\acro{VSBPP}{Variable Size Bin Packing Problem}
\acro{UART}{Universal Asynchronous Receiver/Transmitter}
\acro{SDN}{Software-defined networking}
\acro{CPS}{Cyber Physical System}
\acro{SFC}{Service Function Chaining}
\acro{FSK}{Frequency-shift keying}
\acro{DBPSK}{differential binary phase-shift keying}
\acro{GFSK}{Gausian FSK}
\acro{AES}{Advanced Encryption Standard}
\acro{3GPP}{3rd Generation Partnership Project}
\acro{LTE}{Long-Term Evolution}
\acro{DSL}{digital subscriber line}
\end{acronym}

\usepackage{array}
\newcolumntype{L}[1]{
  >{\raggedright\let\newline\\\arraybackslash\hspace{0pt}}m{#1}}
\newcolumntype{C}[1]{
  >{\centering\let\newline\\\arraybackslash\hspace{0pt}}m{#1}}
\newcolumntype{R}[1]{
  >{\raggedleft\let\newline\\\arraybackslash\hspace{0pt}}m{#1}}

\pgfplotsset{compat=1.17} 

\begin{document}
%
\title{\system: Can we achieve Network Resiliency at the IoT Edge using LPWAN and WiFi?}
%
%
%

\author{Vijay~Kumar,
        Poonam~Yadav,~\IEEEmembership{member,~IEEE,}         and~Leandro~Soares~Indrusiak,~\IEEEmembership{member,~IEEE}
\thanks{V. Kumar is  with the Department of Civil Engineering, University of Bristol, UK, e-mail: vijay.kumar@bristol.ac.uk.}
\thanks{P. Yadav and L.S. Indrusiak are with Computer Science Department, University of York, UK, e-mail: poonam.yadav@york.ac.uk and  leandro.indrusiak@york.ac.uk, respectively.}}

%
%

\markboth{Under Review}
{Kumar \MakeLowercase{\textit{et al.}}: \system: Can we achieve Network Resiliency at the IoT Edge using LPWAN and WiFi? }
%



\maketitle

\begin{abstract}
Edge computing has gained attention in recent years due to the adoption of many Internet of Things (IoT) applications in domestic, industrial and wild settings. The resiliency and reliability requirements of these applications vary from non-critical (best delivery efforts) to safety-critical with time-bounded guarantees. The network connectivity of IoT edge devices remains the central critical component that needs to meet the time-bounded \ac{QoS} and fault-tolerance guarantees of the applications. Therefore, in this work, we systematically investigate how to meet IoT applications mixed-criticality QoS requirements in multi-communication networks. We (i) present the network resiliency requirements of IoT applications by defining a system model (ii) analyse and evaluate the bandwidth, latency, throughput, maximum packet size of many state-of-the-art LPWAN technologies, such as \sigfox, \lora, and \lte and \wifi, (iii) implement and evaluate an adaptive system \system and Criticality-Aware Best Fit (CABF) resource allocation algorithm to meet the application resiliency requirements using Raspberry Pi 4 and Pycom \fipy development board having five multi-communication networks. We present our findings on how to achieve 100\% of the best-effort high criticality level message delivery using multi-communication networks.


\end{abstract}


%
\IEEEpeerreviewmaketitle

\section{Introduction}
\noindent
\IEEEPARstart{I}{oT} devices are everywhere sensing, collecting data and providing information to make a better-informed decision about the environment. Many safety-critical IoT applications such as self-health monitoring through wearable IoT devices connect to a mobile phone/local hub via \bluetooth, \zigbee or \wifi and further send the data to a cloud service or hospital central processing system through Internet~\cite{PPPTestbed2016,Raymond2012}.


In the event of a network-failure, e.g.,  power outage or any other incidental connection failures, the \wifi could be disconnected temporarily, resulting in either data loss or delayed data communication~\cite{Yadav2019}. Depending on the time of the day, it may take from one minute to several minutes to regain connectivity to the \wifi. On the other hand, according to a survey~\cite{BristolBroadband}, the average amount of broadband downtime per year in the UK ranges from 25.4 hours to 168.9 hours.
However, for safety-critical applications, it is essential to maintain resilient data connectivity at all time for the delivery of a time-critical message. The LPWAN technologies have explicitly been designed to meet IoT application requirements. They are built on existing cellular systems to provide improved battery life, power efficiency and indoor and outdoor coverage area~\cite{Rubio-Aparicio2019} at an affordable cost. 
Availability of alternate low power long-range network mediums at a meagre cost opens a new horizon of opportunities.

However, LPWAN technologies also have challenges in terms of limited bandwidth; the number of messages allowed per day and payload size. On the other hand, IoT edge application requirements are defined in terms of message criticality (such as high/low priority), privacy settings, message data length, message sending frequency and user trust on a particular network. Based on the application requirements and available network medium, application traffic can be routed through a specific network medium. Further, in case of a particular network medium unavailability or failure, the application can be informed of the network state and can decide on the suitability of the network and adapt accordingly. For instance, assuming the application is sending data over \wifi and because of power failure \wifi is disconnected, the application can choose to send data over \ac{LTE}(LTE for Machines (LTE-M)/NarrowBand-IoT (\nbiot)), \lora (Long Range), \sigfox and adapt parameters such as payload size and frequency accordingly. 
Despite all the hype and hope of LPWAN, it is not fully understood that if we can achieve network resiliency at the Edge using LPWAN and \wifi for time-critical IoT applications. 
Therefore in this work, we propose a hypothesis that using LPWAN  technologies and \wifi, we can achieve network resiliency at the edge IoT device by providing a capability to choose a suitable network medium based on the application requirements. For the implementation, we utilise affordable, readily-available MicroPython enabled, multi-network micro-controller Pycom \fipy board~\cite{Fipy2020} providing connectivity to Bluetooth, \wifi, \lora, \lte and \sigfox.\\

\noindgras{Contributions:}
\begin{itemize}
    \item We present use cases for resiliency requirements of the IoT edge networks;
    \item provide a detailed analysis  of many state-of-the-art LPWAN technologies, such as \sigfox, \lora, \lte and evaluate their bandwidth, latency, throughput and maximum packet size using an experiment;
    \item define an adaptive system \textit{\system} to meet the application resiliency requirements using underlying LPWAN technologies;
    \item provide open-source implementation of \textit{\system} and detailed insights considering hardware and network limitations.
\end{itemize}

The remainder of this paper is organised as follows:
\autoref{sec:background} provides a technical background about the different LPWAN technologies. In \autoref{sec:system-model}, we define the adaptive \system to meet the application resiliency requirements by providing   two example applications.
In \autoref{sec:resourcemanagement}, we formulate a criticality-aware QoS allocation problem using \ac{ILP} and bin packing algorithms. \autoref{sec:implementation} provides the implementation details of \system prototype and evaluate the baseline metrics. In \autoref{sec:evaluation}, we perform the evaluation of our prototype and discuss hardware and network limitations. In \autoref{sec:related_work} and \autoref{sec:conclusion}, we present related work and conclusion, respectively.

\section{Background}
\label{sec:background}
In this section, we provide a background on multi-mode communication network technologies such as LPWAN technologies (\lora, \sigfox, \lte) and \wifi that we use to provide resilience through redundancy in the \system end-to-end system as shown in  Figure~\ref{fig:block}. We provide a brief introduction to the technology, its range, use-case, security and energy-efficiency. We also provide various performance metrics (max payload length, the possibility of sending continuous data, latency, throughput,  time to connect and reconnect) stated and observed in the wild in \autoref{subsec:evaluation_metrics}. 
\subsection{LPWAN Technologies}
\noindent

\subsubsection*{\textbf{\lora}}
\lora is an \ac{RF} modulation technology for low-power, wide area networks (LPWANs) protocol developed by Semtech. It has a range of up to 5 KM in urban areas and up to 15 KM or more in rural areas (line of sight)~\cite{Semtech2020}. \lora is suitable for specific use cases having requirements of long-range, low power, low cost, low bandwidth, secure with coverage everywhere. For example, measuring water flow using a water flow meter~\cite{lora_water} sending data over \lora. 

A \lora based network consists of end devices, gateways, a network server, and application servers. End devices send data to gateways (\ac{UL}) using single-hop \lora or \ac{FSK} communication. The gateways send the data to the network server via a secured \ac{IP} connection, which, in turn, passes it on to the application server. Additionally, the network server can send messages (either for network management or on behalf of the application server) through the gateways to the end devices (\ac{DL}). \lora allows intermediate gateways to relay messages between the end-devices to the network server, which routes it to the associated application server. Communication between the end-devices and gateway is performed on different frequencies and data rates, which is a trade-off between message length, communication range~\cite{LoRaAllianceTechnicalCommitee2017}. The data transfer from the end device to the application server is encrypted using \ac{AES}~\cite{MEKKI20191}.

From the energy-efficiency perspective, \lora devices have three classes~\cite{Workgroup2015}. Class A device can send data anytime and opens two receive windows after one and two seconds after an \ac{UL} transmission. They are the most energy-efficient; however, the \ac{DL} is only available after transmission. Class B is energy efficient with latency controlled \ac{DL}. They utilize time-synchronized beacons transmitted by the gateway to sync up receive windows. Class C is not efficient in terms of power as they keep the receive window open after transmission~\cite{lora_classA}. \lora also implements \ac{ADR} by managing the data rate and \ac{RF} output for each end-device individually to maximize battery life and maintain network capacity.

\subsubsection*{\textbf{\sigfox}}
Sigfox uses publicly available and unlicensed bands to exchange radio messages over the air (868-869 MHz and 902-928 MHz). It uses \ac{UNB} technology combined with \ac{DBPSK} and \ac{GFSK} modulation. It has a range of approximately 10 km (urban), 40 km (rural). \sigfox mainly caters to IoT applications allowing small messages. For example, a letterbox sensor~\cite{Sigfox_letterbox} sending a message to the user on receiving a post.

The end-device sends the message to the base stations (gateways), which forwards it to the \sigfox backend via a backhaul (3G/4G/\ac{DSL}/Satellite). The backend stores the messages to be retrieved by the end-user via browser/\ac{REST} \ac{API} or set up a callback. For achieving high \ac{QoS}, the end-device sends the message at a random frequency and then sends two replicas on different frequency and time (time and frequency diversity). The message can be received by any number of base stations (spatial diversity). However, \sigfox does not provide any authentication or encryption for the message and device~\cite{MEKKI20191}.

From an energy-saving perspective, the end-device does not require pairing or sending synchronization messages to send the message, thus increasing battery life~\cite{Sigfox2017a}.

\subsubsection*{\textbf{\lte}}
\nbiot is a \ac{3GPP} radio technology standard designed for extended range operation, higher deployment density, and in-building penetration. It utilizes 180 kHz bandwidth and is deployed in-band, guard-band, or standalone mode. On the other hand, LTE-M provides high latency communication, support for extended coverage, LTE-M half-duplex mode/full-duplex, \ac{SMS}, coverage enhancement, connected mode mobility~\cite{GSMA2018}. Both NB-IoT and LTE-M have a range of approx. 1 km (urban) and 10 km (rural)~\cite{MEKKI20191}. Both follow 3GPP standards and have LTE encryption by default.

NB-IoT is suited for static, low throughput, and low power applications. For example, Nortrace tracked sheep's location and well-being over mountainous regions using NB-IoT~\cite{NB_IoT_sheep}. In contrast, LTE-M is best for applications requiring mobility, voice, and \ac{SMS}~\cite{Arkessa20}. For example, Telstra tracks the location of high-value, non-powered assets, such as shipping containers, semi-trailers, rail freight wagons, and large machinery using LTE-M~\cite{GSMAssociation2019}.

From the energy-efficiency perspective, both include \ac{PSM} and \ac{eDRX}. PSM reduces the energy used by \ac{UE} which defines how often and how long the \ac{UE} will be active to send and receive data. \ac{eDRX} improve end-device life for mobile-terminated traffic by switching off the receiver circuit for a defined period~\cite{GSMA2019}.

\subsubsection*{\textbf{Integration - \lora/\sigfox}}

\label{subsubsec:integrations}

\nbiot/LTE-M is an \ac{IP}-based network allowing data to be transferred to its associated cloud server. However, in the case of the \lora and \sigfox, data is sent to \ac{TTN} console and \sigfox backend, respectively. Currently, \ac{TTN} console and \sigfox backend provide multiple integration methods to retrieve data such as AWS IoT, AllThingsTalk, Microsoft Azure IoT Hub, HTTP, Emails, and other callbacks. 

\subsection{\wifi}
Mostly IoT devices have a low-cost, low-power system on a chip micro-controller with integrated \wifi and dual-mode Bluetooth. \wifi on \iot devices support different wireless modes such as 802.11 b/g/n/e/i,  provide automatic beacon monitoring and scanning. The Pycom \fipy board used in our prototype has a \wifi radio system on chip with $1KM$ \wifi range.

\section{System Model and Motivating Example}
\label{sec:system-model}
\begin{figure}
  \centering
  \includegraphics[width=0.9\columnwidth]{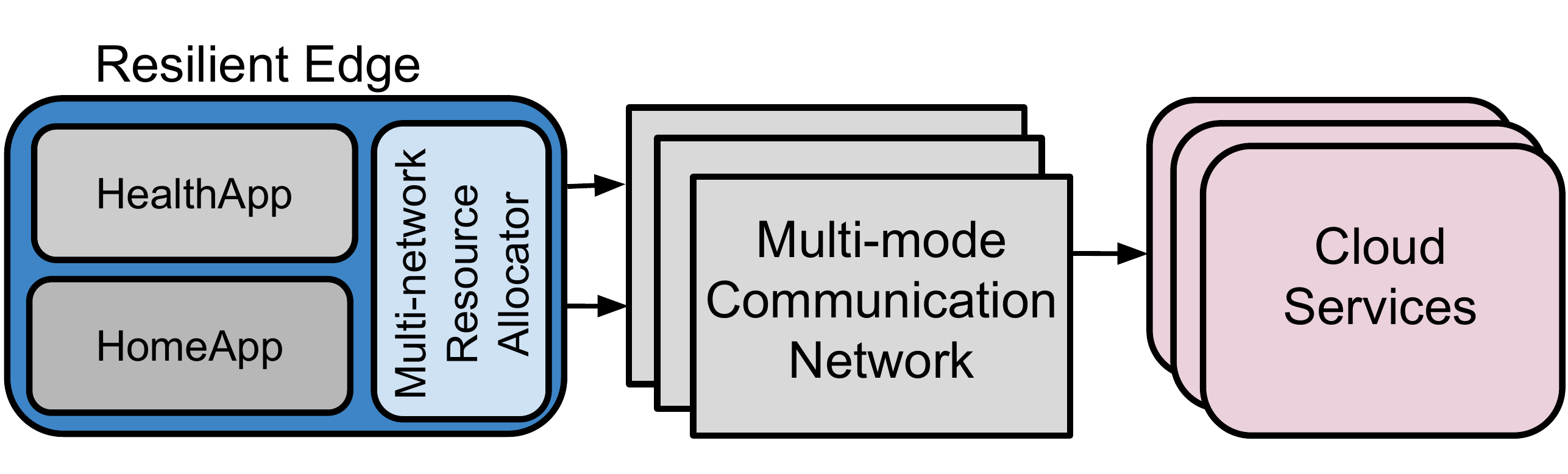}
  
  \caption{Resilient Edge End-to-end System.}
  \label{fig:block}
\end{figure}

To better understand the network requirements of \system applications, we start by considering a use-case with a concrete example. Figure~\ref{fig:block} and \ref{fig:exp_setup} show an edge device running two sample applications to support assisted living facilities: one of the applications monitors the health of the resident (\appX), the other monitors their residential unit (\appY).
In the real-world setting, the similar new applications can be configured for the data rates defined by the application QoS requirements and maximum network bandwidth availability. To achieve continuous network connectivity needed by these applications, we make use of a multi-mode communication network (details are provided in the next section).

\begin{figure*}
  \centering
  \includegraphics[scale=0.9]{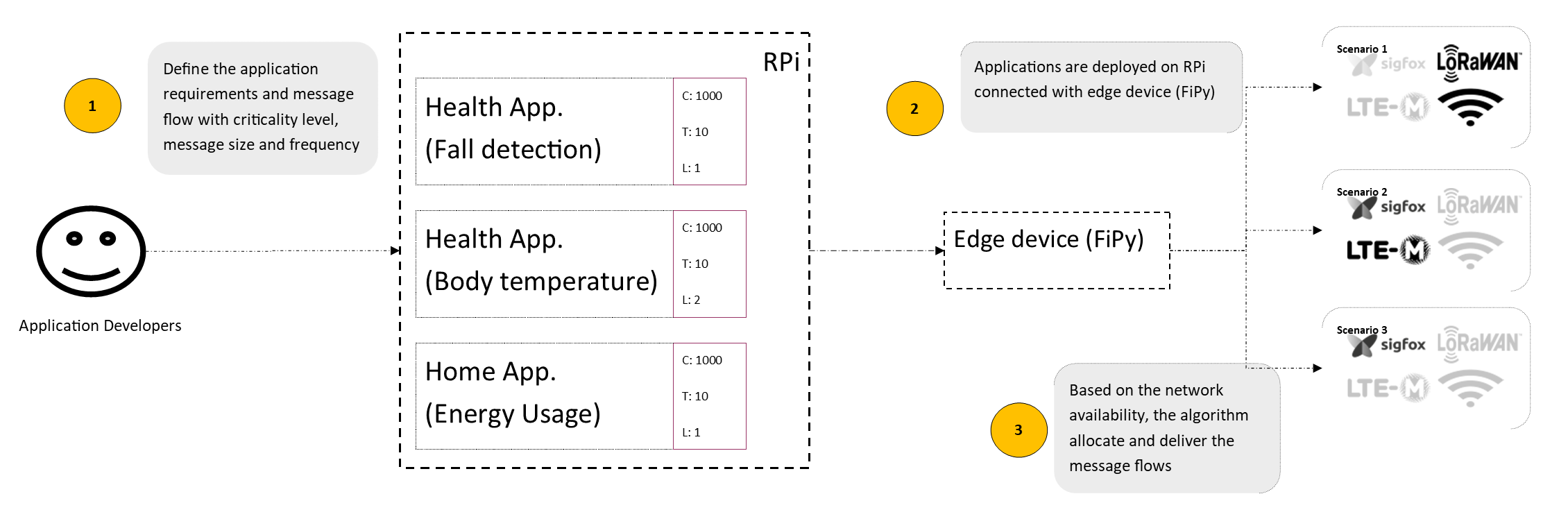}  
  \caption{System model summary with different applications with criticality, message size and frequency defined by application developers and different network availability scenarios (Faded symbol represents network unavailability). }
  \label{fig:exp_setup}
\end{figure*}

We now present an abstract system model defining the attributes of application data flows so that we can reason about the \ac{QoS} needs of each application, and about ways to (partially) fulfill those needs under different scenarios and different levels of multi-network connectivity.
We propose that the communication needs of specific applications must be explicitly declared as message flows. An application can declare an arbitrary number of message flows, and each message flow represents a potentially infinite series of messages to be sent through one of the local network interfaces. To allow application developers to quantitatively declare the \ac{QoS} needs for each message flow, we revisit the notion of mixed-criticality communication proposed in~\cite{AirTight2020} and support the definition of \ac{QoS} requirements at distinct levels of criticality. As in~\cite{AirTight2020}, our goal is to allow the system to guarantee a predefined level of service for all message flows during normal operation, but also provide graceful degradation of service in adverse circumstances by allowing the most critical communication to be maintained. Unlike~\cite{AirTight2020}, however, we are not interested in meeting hard real-time deadlines and will instead use the notion of criticality-specific \ac{QoS} requirements to manage multi-network resources.  

\begin{table}

\makebox[0.48 \textwidth][c]{       
\resizebox{0.48 \textwidth}{!}{   

\centering

\begin{tabular}{c|l|c|c|c|c|c|c|}
\cline{2-8}
\multicolumn{1}{l|}{\textbf{}}             & \multicolumn{1}{r|} {\textbf{Criticality Level}} & \multicolumn{2}{c|}{{\textbf{1}}} & \multicolumn{2}{c|}{\textbf{2}} & \multicolumn{2}{c|}{\textbf{3}}     \\ \hline
\multicolumn{1}{|l|}{\textbf{Applications}} & \textbf{Message Flow $\tau$}           & \textbf{C } & \textbf{T } & \textbf{C} & \textbf{T } & \textbf{C } & \textbf{T } \\ \hline
\multicolumn{1}{|l|}{\appX}    & 1 fall detection                  & 1000               & 10             & 40                 & 20            & 10                 & 60           \\ \hline
\multicolumn{1}{|c|}{"}                    & 2 heart monitoring            & 1000               & 5              & 80                 & 10             & 10                 & 20           \\ \hline
\multicolumn{1}{|c|}{"}                    & 3 body temperature                & 30                 & 30             & 10                 & 120            &                    &                \\ \hline
\multicolumn{1}{|l|}{\appY}      & 4 sensor bedroom                  & 40000              & 10             & 10                 & 30             &                    &                \\ \hline
\multicolumn{1}{|c|}{"}                    & 5 sensor bathroom                 & 80                 & 10             & 10                 & 30             &                    &                \\ \hline
\multicolumn{1}{|c|}{"}                    & 6 sensor lounge/kitchen           & 40000              & 10             & 10                 & 30             &                    &                \\ \hline
\multicolumn{1}{|c|}{"}                    & 7 sensor front door               & 40000              & 10             & 10                 & 30             &                    &                \\ \hline
\multicolumn{1}{|c|}{"}                    & 8 energy usage               & 40              & 3600             &                  &              &                    &                \\ \hline
\end{tabular}
} 
} 
\caption{Message flows on an edge device for assisted living facilities - C is maximum message size (bytes) and T is minimum interval between subsequent message (seconds). We assume that high criticality level messages are necessary to be delivered messages and are rare and have smaller size. In this example, for message flow 1, when criticality level 3 is requested and served, underlying network interface guarantees a message delivery service with message size of 10 bytes with 60 seconds subsequent message interval.}
\label{tab:AssistedLivingExample}
\end{table}

Our model allows system designers and administrators to decide how many levels of criticality $L=L_{max}$ to support, and then to allow the specification of the \ac{QoS} requirements of each message flow at each of those levels.  The \system, as shown in Figure~\ref{fig:block} is designed to support three levels of criticality, and the Table~\ref{tab:AssistedLivingExample} shows the \ac{QoS} required by each message flow at each level. The message flows in Table~\ref{tab:AssistedLivingExample} have been defined by taking a bottom to top approach. The values reflect the prototype application data size and are intuitively set by authors considering several state-of-the-art IoT applications. The message flows and the requirements are driven by the network capacity available on the edge device (FiPy). In reality, the applications (designed by application developers) can request any message size and message interval. The algorithm will try to allocate and deliver the message flows as per the available network capacity.

We assume $L=1$ as the criticality level denoting normal operation mode, so the \ac{QoS} requirements at that level should declare the largest communication volumes and injection rates of each message flow to account for all critical and non-critical traffic. \ac{QoS} requirements at higher levels of criticality should only be declared for message flows that carry critical data and should account only for the necessary communication volumes and injection rates at each of those levels. By declaring or not a \ac{QoS} requirement at a given level, application developers can explicitly distinguish the criticality of each message flow, and to explicitly define a number of service degradation levels each of them can support.

We can now define a message flow $\tau_i$ as a tuple ($A_i$, $C_i$, $T_i$) where $A_i$ denotes the application to which the message flow belongs to, $C_i$ denotes the maximum message size (in bytes) and $T_i$ denotes the minimum interval between subsequent messages of the flow (in seconds). The bandwidth utilisation $U_i$ of a flow $\tau_i$ can be calculated by the quotient $C_i/T_i$. 

To support multiple criticality levels, $C_i$ and $T_i$ are defined as arrays of length $L_{max}$, so $C_i^L$ and $T_i^L$ denote, respectively, the maximum message size and the minimum interval between subsequent messages of $\tau_i$ at criticality level $L$.

 In normal operation (i.e. $L=1$), message flows declare their most generous \ac{QoS} requirements, with larger data volumes for home monitoring (e.g. including camera snapshots in most of them) and resident monitoring (e.g. detail accelerometer data for fall detection, full (electrocardiogram) ECG data for heartbeat monitoring). The next criticality level (i.e. $L=2$) allows the declaration of degraded \ac{QoS} levels, which in this example is provided for all message flows except for the one monitoring energy usage (which will not be forwarded by the edge device in case of degraded service). Notice that the \ac{QoS} requirements declared for $L=2$ and show that monitoring will be performed less often and less data will be provided (e.g. simple movement detectors for home monitoring, average temperature and heartbeat for health monitoring). Finally, only two message flows declare \ac{QoS} requirements at the highest level of criticality (i.e. $L=3$), representing the alarms for fall or severe arrhythmia/cardiac arrest. In the case of degraded service, all available resources should be used to provide those two flows with their declared \ac{QoS} requirements.

\section{Multi-Network Resource Management}
\label{sec:resourcemanagement}
Given the system model proposed in Section \ref{sec:system-model}, we can now formulate a criticality-aware QoS allocation problem.

A straightforward way to ensure QoS to the application message flows is to prevent the over-utilisation of the network interfaces they are assigned to. For example, by providing criticality level $L=2$ guarantees to all message flows of the \appX application from Table \ref{tab:AssistedLivingExample} it would be possible to allocate all of them to a \lora network (as their compound bandwidth utilisation would not exceed 6 bps), but the same network would be over-utilised if flows operate at criticality level $L=1$ (where their compound bandwidth utilisation would exceed 1700 bps).

We can therefore formulate the criticality-aware QoS allocation problem as the choice, for each message flow of each application, of its allowed criticality level of service and its allocated network interface. Such problem is similar to a \ac{VSBPP}~\cite{VSBPP2008}, but with a fixed number of bins (i.e. the different networks, each of them with their bandwidth and payload size limitations) and with a choice of sizes for each element (i.e. the message flows, with their choice of criticality level).

\subsection{ILP Formulation}
\label{sec:ilp}

Similarly to the standard \ac{VSBPP}, we can formulate our problem with an \ac{ILP} model. For the sake of simplicity, we describe the size of bins and elements by their bandwidth capacity and utilisation, respectively. We claim that an extension to a multi-dimensional formulation (i.e. that can also capture maximum payload sizes, maximum number of daily messages, etc.) is straightforward but left as future work. The assumption is that the QoS requirements of all applications can be satisfied provided there is enough bandwidth of one network or combined bandwidth of multiple networks. However, in practice, the network capacity is limited, and there would be applications whose QoS requirements cannot be satisfied.

Let us then consider a set $\mathcal{T}$ of elements representing our message flows $\tau_i, i=1...n$, each of them with a potential choice of values $U^L_i$ representing the different bandwidth utilisations $C^L_i/T^L_i$ at each level of criticality they are designed to support (or $\infty$ if that flow does not specify service at a given criticality level, e.g. \emph{energy usage} flow at levels $L=2$ and $L=3$ in Table \ref{tab:AssistedLivingExample}).

Likewise, let us consider a set $\Gamma$ of bins representing our network interfaces $\gamma_j, j=1...m$, each of them with a bandwidth capacity $B_j$. Finally, we define a set of binary variables $x_{i,j,L} \in \{0,1\}$, and assume that $x_{i,j,L}=1$ if message flow $\tau_i$ is assigned to network $\gamma_j$ and configured to operate at criticality level $L$, or $x_{i,j,L}=0$ otherwise. Given the ranges ${1}\le{i}\le{n},~1\le{j}\le{m},~1\le L\le{L_{max}}$ we will have at most $n \times m \times L_{max}$ binary variables for a given problem.

To ensure the assignment of values to the binary variables represent a valid solution to our problem, we must now state a number of constraints. First, we make sure that a message flow $\tau_i$ is allocated to a single network interface and configured to operate at a single criticality level by stating that $\sum\limits_{j=1}^m {\sum\limits_{L=1}^{L_{max}} x_{i,j,L}} = 1$ for all $1 \le i \le n$. Secondly, we ensure that no network interface $\gamma_j$ is overloaded by stating that $\sum\limits_{i=1}^n {\sum\limits_{L=1}^{L_{max}} x_{i,j,L} \times U^L_i} \le B_j$ for all  $1 \le j \le m$.

Finally, we can state our maximisation objective function as: 
\begin{equation}
\label{eq:ObjectiveFunction}
objective = \sum\limits_{i=1}^n {\sum\limits_{j=1}^m {\sum\limits_{L=1}^{L_{max}} x_{i,j,L} \times (1+L_{max}-L)}}
\end{equation}

The rationale behind the maximisation of the objective is to configure message flows at the lowest possible levels of criticality (i.e. lowest values for $L$), thus providing each message flow with the most generous possible QoS, while avoiding network overload. The unit added to the last term of the equation is crucial to allow the objective to distinguish a flow that is allocated at the highest criticality and one that is not allocated at all. 

An additional constraint could be formulated, in case all message flows must be allocated to a network interface and receive some level of service: $\sum\limits_{i=1}^n {\sum\limits_{j=1}^m {\sum\limits_{L=1}^{L_{max}} x_{i,j,L}}} = n$. This is not always necessary or desirable, as it may the intention of application designers that, under limited network availability, only a subset of the application message flows should be provided service (e.g. in the example from Table \ref{tab:AssistedLivingExample}, under the most stringent conditions at $L=3$, only the \emph{fall detection} and \emph{heart monitoring} message flows require service). In such cases, such a constraint may be rewritten to ensure that specific message flows are always allocated service, or even be reformulated as part of the objective function, aiming to maximise the number of message flows that are guaranteed some level of service.

\subsection{Bin-Packing Algorithms}

While the formulation given in subsection \ref{sec:ilp} can be optimally solved by an ILP solver, it may not be reasonable to expect that such a software package could be installed and executed by resource-constrained edge devices such as the ones considered in this work. We, therefore, propose the use of simple bin-packing algorithms that are able to achieve acceptable results with a much lower computational overhead. In particular, we define a criticality-aware best fit (CABF) algorithm and show its performance compared to classic first fit, best fit, and worst fit algorithms (FF, BF, and WF) as well as their decreasing variants (FFD, BFD, and WFD). 

Since the classic algorithms are unaware of the different criticality levels, we implemented two alternatives for each of them, one that tries to fit message flows to networks at their highest level of criticality (i.e. H-FF, H-BF, H-WF and their decreasing counterparts) and another that does the same with the lowest defined criticality of each message flow (i.e. L-FF, L-BF, L-WF and their decreasing counterparts).

Algorithm \ref{alg:CABF} describes the proposed CABF algorithm, which takes as inputs the sets $\mathcal{T}$ of message flows and $\Gamma$ of networks, and outputs a set $Q$ of 3-tuples $q=(\tau_i,\gamma_j,L)$, each of them representing the allocation of a message flow $\tau_i$ to a network $\gamma_j$ at criticality level $L$. Algorithm \ref{alg:CABF} uses the following notation: $q(\tau)$, $q(\gamma)$ and $q(L)$ denote the first, second and third element of a 3-tuple $q$, and likewise $Q(\tau)$, $Q(\gamma)$ and $Q(L)$ denote the sets of all first, second and third element of the 3-tuples in $Q$; $\mathcal{T}_L$ is the subset of $\mathcal{T}$ including all message flows that are declared at a given criticality level $L$ (as not all flows must be declared for all levels); and $BestFit(\tau, L, \Gamma)$ denotes a function which returns the network $\gamma \in \Gamma$ which is the best fit allocation for the message flow $\tau$ at its criticality level $L$, or $\emptyset$ if $\tau$ does not fit in any of the networks in $\Gamma$, taking into account the allocations already in $Q$.

\begin{algorithm}
\caption{Criticality-Aware Best Fit (CABF)}
\label{alg:CABF}
\KwResult{Set of 3-tuples indicating the allocated network and configured criticality level for all message flows that can be provided service}

  \SetKwInOut{Input}{inputs}
  \SetKwInOut{Output}{output}
  \SetKwProg{CABF}{CABF}{}{\textbf{end}}

 \CABF{$(\mathcal{T},\Gamma)$}{
    \Input{set $\mathcal{T}$ of message flows, set $\Gamma$ of networks}
    \Output{set $Q$ of 3-tuples $q=(\tau_i,\gamma_j,L)$}
     $Q \gets \emptyset$\;
     \For{$(l = L_{max};\ l > 0;\ l=l-1)$} {
        
        \ForEach{$(q \in Q \mid q(\tau)\in \mathcal{T}_l \land q(L)>l)$}{
            $Q \gets Q - q$\;
            $\gamma_{reloc} \gets BestFit(q(\tau), l, \Gamma)$\;
            \If{$\gamma_{reloc} \ne \emptyset$}{ 
                $q \gets (\tau,\gamma_{reloc},l)$\;
            }
            $Q \gets Q + q$\;
        }
        
        \ForEach{$(\tau_{new} \in \mathcal{T}_l \mid \tau_{new} \notin Q(\tau))$}{
        
            $\gamma_{new} \gets BestFit(\tau_{new}, l, \Gamma)$\;
            \If{$\gamma_{new} \ne \emptyset$}{ 
                $Q \gets Q + (\tau_{new},\gamma_{new},l)$\;
            }
            
        }
     }
     \KwRet{$Q$}\;
 }
\end{algorithm}

The intuition behind the CABF algorithm is as follows. It tries to allocate first the message flows defined at higher criticalities, as shown by the outer \emph{for} loop decreasing from $L_{max}$ to 1. As it iterates over that loop towards lower criticality levels, and before it allocates flows defined at the criticality level of the current iteration, it first attempts to lower the criticalities of flows allocated in the previous iterations. This is shown by the first inner \emph{forall} loop, which iterates over tuples in $Q$ with flows that have definitions at the criticality level of the current iteration (i.e. $q(\tau)\in \mathcal{T}_l$). Within the first inner \emph{forall} loop, the algorithm removes the original allocation from $Q$, then tries to find a network $\gamma_{reloc}$ which is the best fit for the flow using its lower criticality figures. If the best-fit algorithm succeeds to find an allocation with the lower criticality values, a 3-tuple representing that new allocation is added to $Q$. If it fails to find a network that is able to accommodate the requirements at a lower criticality, the original allocation is returned back to $Q$. Once the first inner \emph{forall} loop finishes, the second inner \emph{forall} loop uses the best-fit algorithm to allocate, when possible, all unallocated message flows that have definitions at the criticality level of the current iteration.

The proposed order of the two inner \emph{forall} loops reflects an assumption that flows that have definitions at higher levels of criticality should always be given more resources if they become available. This will not always be the case in every application domain, and in many cases it may be better to first use resources to provide some service to less-critical message flows rather than improve the service to highly-critical ones. Reversing the proposed order of the two inner \emph{forall} loops would achieve exactly that, therefore we name that variant $CABF_{inv}$.

\subsection{Evaluation}

Table \ref{tab:AssistedLivingExampleResults} shows the network allocations and choice of criticality level for each of the message flows of the motivating example described in Section \ref{sec:system-model}. The table shows allocations produced by each of the baseline bin-packing algorithms, by both variants of the proposed algorithm, and one solution (out of many possible ones) produced by an optimal solver. The allocations assume the availability of three networks with bandwidths of 64000, 1760 and 48 bits per second, representing \wifi, \lora SF9 and \sigfox networks (but disregarding maximum payload size or the number of daily messages), and represented by the symbols $*$, $\#$ and $+$, respectively. 

Both variations of the proposed algorithm are able to produce optimal solutions in this example, providing service to all flows, with all-but-two at their lowest criticality level (which leads to an objective result of 22 according to Equation \ref{eq:ObjectiveFunction}). 

\begin{table}[h]
\makebox[0.48 \textwidth][c]{       
\resizebox{0.48 \textwidth}{!}{   
\renewcommand{\arraystretch}{1}
\begin{tabular}{c|c|c|c|c|c|c|c|c|c|c|c|}
\cline{2-12}
\textbf{} & \multicolumn{8}{c|}{\textbf{Message Flows}} &
\multirow{1}{*}{\textbf{\begin{tabular}[c]{@{}c@{}}\% flows\end{tabular}}} &
\multirow{1}{*}{\textbf{\begin{tabular}[c]{@{}c@{}}avg \\\end{tabular}}} & \multirow{1}{*}{\textbf{objective}} \\ \cline{2-9}
    & 1 & 2 & 3 & 4 & 5    & 6    & 7    & 8    &  \textbf{served}  &  \textbf{crit} &
     \\ \cline{1-9}
\multicolumn{1}{|c|}{\textbf{\begin{tabular}[c]{@{}c@{}}Requested \\ crit level\end{tabular}}} & 1,2,3  & 1,2,3  & 1,2  & 1,2  & 1,2  & 1,2  & 1,2  & 1    &   & \textbf{level}    &                                     \\ \cline{1-9}
\multicolumn{1}{|c|}{\textbf{\begin{tabular}[c]{@{}l@{}}Allocation\\ algorithms\end{tabular}}}       & \multicolumn{8}{c|}{\textbf{Allocated Criticality Level}}                                     &                                          &                                                 &                                     \\ \hline
\multicolumn{1}{|l|}{L-FF}                                                                     & 1*         & 1*         & 1*         & 1*         & 1*         &            &            & 1*         & 75                                                                                   & 1                                                                                            & 18                                  \\ \hline
\multicolumn{1}{|l|}{L-FFD}                                                                    &            & 1\#       & 1\#        & 1*         & 1\#        & 1*         &            & 1+         & 75                                                                                   & 1                                                                                            & 18                                  \\ \hline
\multicolumn{1}{|l|}{H-FF}                                                                     & 3*         & 3*         & 2*         & 2*         & 2*         & 2*         & 2*         & 1*         & 100                                                                                  & 2.12                                                                                         & 15                                  \\ \hline
\multicolumn{1}{|l|}{H-FFD}                                                                    & 3*         & 3*         & 2*         & 2*         & 2*         & 2*         & 2*         & 1*         & 100                                                                                  & 2.12                                                                                         & 15                                  \\ \hline
\multicolumn{1}{|l|}{L-WF}                                                                     & 1*         & 1*         & 1*         & 1*         & 1*         &            &            & 1*         & 75                                                                                   & 1                                                                                            & 18                                  \\ \hline
\multicolumn{1}{|l|}{L-WFD}                                                                    &            & 1\#       & 1\#        & 1*         & 1\#        & 1*         &            & 1+         & 75                                                                                   & 1                                                                                            & 18                                  \\ \hline
\multicolumn{1}{|l|}{H-WF}                                                                     & 3*         & 3*         & 2*         & 2*         & 2*         & 2*         & 2*         & 1*         & 100                                                                                  & 2.12                                                                                         & 15                                  \\ \hline
\multicolumn{1}{|l|}{H-WFD}                                                                    & 3*         & 3*         & 2*         & 2*         & 2*         & 2*         & 2*         & 1*         & 100                                                                                  & 2.12                                                                                         & 15                                  \\ \hline
\multicolumn{1}{|l|}{L-BF}                                                                     & 1\#        & 1*         & 1+         & 1*         & 1\#        &            &            & 1+         & 75                                                                                   & 1                                                                                            & 18                                  \\ \hline
\multicolumn{1}{|l|}{L-BFD}                                                                    &            & 1\#       & 1+         & 1*         & 1\#        & 1*         &            & 1+         & 75                                                                                   & 1                                                                                            & 18                                  \\ \hline
\multicolumn{1}{|l|}{H-BF}                                                                     & 3+         & 3+         & 2+         & 2+         & 2+         & 2+         & 2+         & 1+         & 100                                                                                  & 2.12                                                                                         & 15                                  \\ \hline
\multicolumn{1}{|l|}{H-BFD}                                                                    & 3+         & 3+         & 2+         & 2+         & 2+         & 2+         & 2+         & 1+         & 100                                                                                  & 2.12                                                                                         & 15                                  \\ \hline
\multicolumn{1}{|l|}{CABF}                                                                     & 2+         & 1\#        & 1+         & 2+         & 1\#        & 1*         & 1*         & 1+         & 100                                                                                  & 1.25                                                                                         & 22                                  \\ \hline
\multicolumn{1}{|l|}{CABFinv}                                                                  & 1\#        & 2\#        & 1+         & 2+         & 1\#        & 1*         & 1*         & 1+         & 100                                                                                  & 1.25                                                                                         & 22                                  \\ \hline
\multicolumn{1}{|l|}{Optimal}                                                                  & 1*         & 1*         & 1*         & 1*         & 1*         & 2*         & 2*         & 1*         & 100                                                                                  & 1.25                                                                                         & 22                                  \\ \hline

\end{tabular}

} 
} 
\caption{Obtained criticality level (1 $|$ 2 $|$ 3) and network allocation (* Wi-Fi $|$ \# Lora $|$ + Sigfox) for motivating example.}
\label{tab:AssistedLivingExampleResults}
\end{table}

\section{Implementation}
\label{sec:implementation}
\begin{figure*}
    \scalebox{.7}{\input{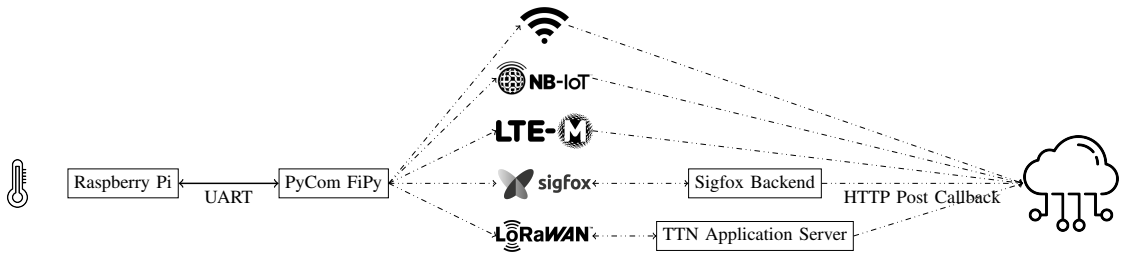}}
    \caption{Block diagram of current experimental setup.}
    \label{fig:exp_block}
\end{figure*}

The \system prototype setup is shown in Figures~\ref{fig:exp_block} and \ref{fig:exp_setup}. A Raspberry Pi model 4~\cite{Rpi2020} (\rpi) is interfaced with Pycom \fipy~\cite{Fipy2020}.
\rpi has Broadcom BCM2711, Quad-core Cortex-A72 (ARM v8) 64-bit SoC @ 1.5GHz with 4 GB RAM, \fipy has an Xtensa® dual-core 32–bit LX6 microprocessor and on-chip SRAM of 520KB and external SRAM 4MB with an external flash of 8 MB. 
\fipy provides connectivity to five different networks \wifi, \bluetooth, \lora, \sigfox and \lte. More details about the interworking of \fipy can be found in the \fipy datasheet~\cite{FiPY_Datasheet}. The \rpi \ac{UART} (GPIO14-TXD/ GPIO15-RXD) is connected to the expansion board pins (P3-TXD/P4-RXD) of \fipy to transfer the data from the \rpi to \fipy. We implemented and simulated the message flow of \appX and \appY on \rpi and multi-network resource allocator on \fipy, respectively.
To enable the data transfer between \rpi and \fipy, a message payload from the applications is written to the \rpi UART and read by \fipy continuously. 
On \fipy, a python script checks the messages received from the \rpi, the network interface assigned to the message flow, its criticality level, and attempt to send it via that network interface.

\begin{figure}
  \centering
  \includegraphics[width=0.8\columnwidth]{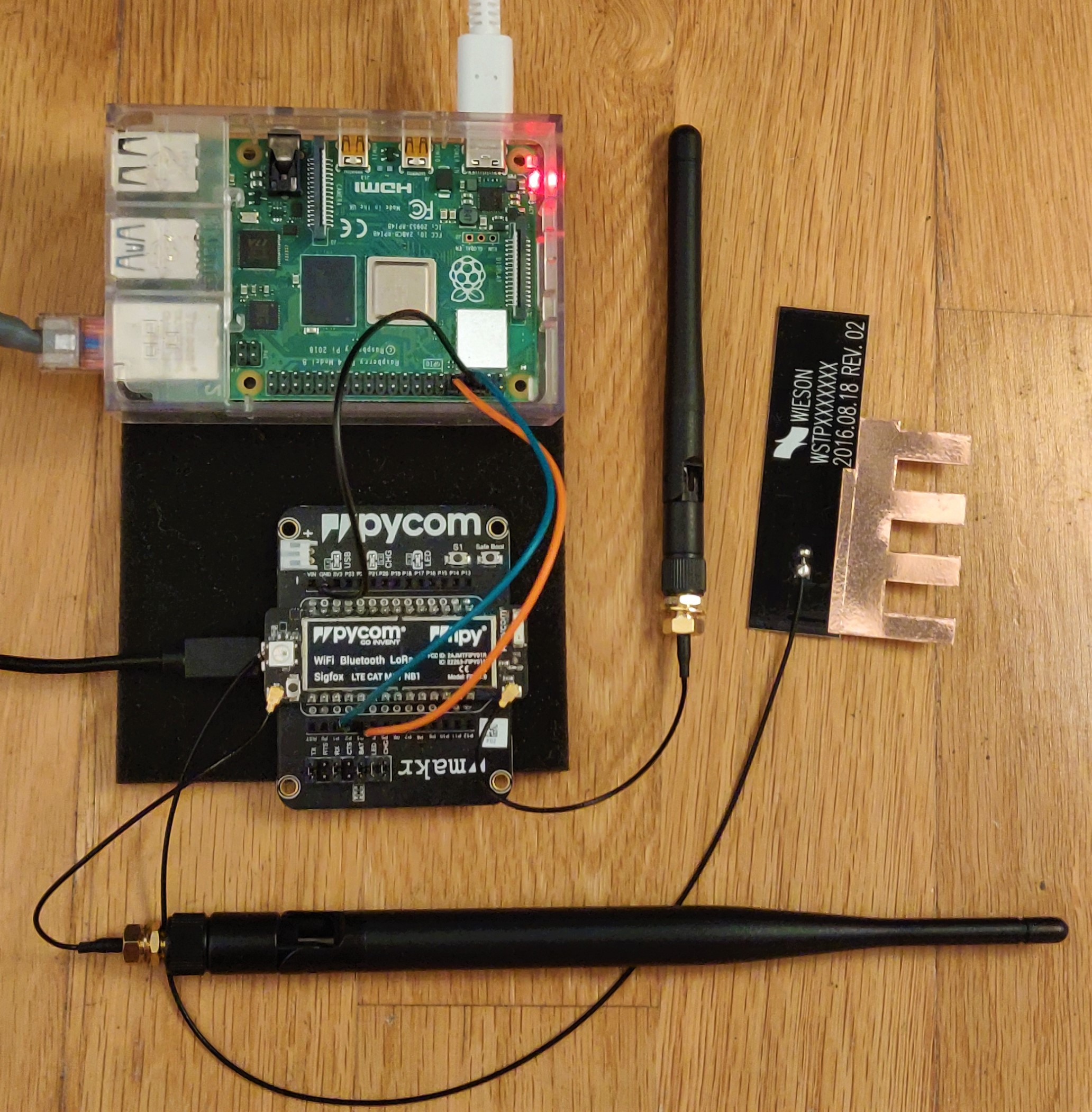}
  \caption{Current Experimental Setup.}
  \label{fig:exp_setup}
\end{figure}




\subsubsection*{\textbf{Implementation details of \rpi components}} We utilise \ac{TCP}/\ac{IP} serial bridge\footnote{TCP/IP - serial bridge \url{https://pyserial.readthedocs.io/en/latest/examples.html\#tcp-ip-serial-bridge}} to create a socket listening on port 8080 connecting to the UART (/dev/ttyAMA0) to send and receive data to and from the UART. Further, we utilise \textit{python select lib}~\footnote{Select - Waiting for I/O completion \url{https://docs.python.org/3/library/select.html}} to monitor sockets for incoming data to be read and send outgoing data when there is room in the buffer and utilise message queues to store the outgoing messages. To send and receive a message over UART efficiently and without breaking, we add a header with \texttt{(:ML:<MessageLength>)} at a start and a newline \verb!'\n'! at the end. On both sides, \rpi reading a socket and \fipy reading the UART, we ensure that we have received the full message. To simulate the message flows running on the \rpi, we utilise threads to write a message payload on the socket with \texttt{<MessageFlow Name, Criticality level, the payload>}. The thread sleeps for the period specified by the message flow before sending the next message. We store the statistics about the number of messages sent by a particular message flow, acknowledgment or error received.

The \fipy runs a multi-network resource allocator and sends an allocation message back to the \rpi stating which message flows have been assigned with which criticality level. A sample Message Flow Element Allocation (MFEA) message is:\\\\
\texttt{MFEA:[{'PS': 41, 'N': 'Wi-Fi', 'PE': 10, 'MF': 'Kitchen Sensor', 'CL': 1}]}\\
where $PS$ is payload size, $N$ is network assigned, $PE$ is the period (time in seconds) between two subsequent messages,  $MF$ is message flow name, and $CL$ is criticality level assigned. If the network conditions at the \fipy are changed, and a new MFEA message is received, the previous thread sending the messages are stopped, and new threads for sending a message with a particular criticality level and period are launched.

\subsubsection*{\textbf{Implementation details of FiPy components}}
On FiPy, before assigning any network to a message flow, we need to create a network \verb!"bin"! of the available networks (\wifi, \lte, \lora and \sigfox) and add the corresponding network interfaces to the network \verb!"bins"!. While doing so, we take into some limitations that are posed by underlying hardware such as if \wifi is available, we do not add a network \verb!"bin"! for \lte because in the current version of \fipy if both \wifi and \lte are connected at the same time, \fipy does not provide routing capabilities to direct the traffic~\cite{FiPy_WiFiNBIoT}. If \wifi is unavailable, then we connect via \lte. Similarly, if the \lora network is available, we add \lora to the network \verb!"bin"!. If \lora is unavailable, we add \sigfox, mainly because \sigfox and \lora share the same radio module.
As part of the Multi-network resource allocator  - we implement variant Criticality-Aware Best Fit ($CABF_{inv}$) and set the initial parameters, and perform the allocations of the message flows to the network interface. After the allocations, we continuously read the UART for the messages from the \rpi. The messages from the \rpi are in the format of \texttt{<MessageFlow Name, Criticality Level, Payload>}. On the \fipy, we check if the Message Flow with criticality level has been assigned; if assigned, an attempt to send the payload is made. If the message flow is not allocated, an error message is sent back to \rpi, mentioning message flow is not allocated. Similarly, if the payload is delivered, an ACK message is sent to the \rpi; if not delivered, an error message with not delivered is sent through UART.

FiPy provides multi-network connectivity, and powering on all the network interfaces could result in significant power consumption. With that in mind, currently, we initialize all the network interfaces at the boot and connect to a specific network based on network availability and conditions. For instance, the NB-IoT connection is skipped if the Wi-Fi network is available; if the LoRa network joins successfully, the Sigfox socket is not created. Further, we have mentioned the time (in seconds) for different technologies to connect to the network (\autoref{subsubsec_time_to_connect}) and time complexity and context switching of the $CABF_{inv}$ algorithm (\autoref{subsec:time_complexity}) that provides a rough estimate on switching overhead if the IoT devices need to switch between network interfaces and to turn on/off the interface.

The multi-network resource meets the QoS requirements of IoT applications by determining the different network interfaces available and the communication parameters of the selected technology (bandwidth). For instance, when a network interface is defined (whether
it is available or not), we determine the bandwidth provided by that network. For instance, \lora starts the connection with adaptive \ac{SF}, i.e., it would start with SF7; if it did not connect with SF7, it would try with SF8 and so on. Based on the connection, we take the bandwidth of the connected SF.

\subsubsection*{\textbf{Receiving messages on Cloud}}
To store the messages sent by the \fipy (as shown in \autoref{fig:exp_block}), we utilise Tornado - a python web framework and asynchronous networking library \footnote{Tornado Web Server \url{https://www.tornadoweb.org/en/stable/}} to run an \texttt{HTTP} server on a machine hosted on a cloud. The \texttt{HTTP} server accepts \texttt{HTTP POST} messages and receives them directly from the \fipy via \wifi, \ac{TTN} application server via \lora, \sigfox backend via \sigfox and Pybytes\footnote{Pybytes \url{https://docs.pycom.io/pybytes/}} via \nbiot. When the message flow allocated interface is \wifi, an \texttt{HTTP POST} request is sent from \fipy to the cloud machine using \texttt{urequests micro-python library}. When the assigned interface for message flow is \lora, \sigfox and \nbiot, the message is sent via the respective interface. On \ac{TTN} application server, \sigfox backend and Pybytes for \nbiot, we have configured the \texttt{HTTP} Integration as defined in~\autoref{subsubsec:integrations}. HTTP integration sends the \ac{UL} data received from \fipy to our cloud machine. The \texttt{HTTP} server checks for the \texttt{URI} and fetch the data from the post data and stores it in a \texttt{influxdb} database.

\subsection{Evaluation of Baseline Metrics}
\label{subsec:evaluation_metrics}
For performance evaluation, we considered the following metrics: max payload length, inter-message gap, latency, throughput, time to connect and reconnect, and performed the initial experiments to get the baseline results for each network (\lora, \sigfox, \nbiot, \wifi) before deploying the multi-network resource allocator on \fipy. In \autoref{tab:baseline_summary2}, we provide a summary of the metrics found in these baseline experiments. Application developers can decide on the suitable network medium for the application based on the application requirements and the use-case. First, we provide how each network compares with each other, followed by more information about the experiment.

\begin{table*}[ht]
\makebox[1\textwidth][c]{       

\begin{tabular}{L{0.2\linewidth}| L{0.18\linewidth}| L{0.09\linewidth}|L{0.25\linewidth}|L{0.14\linewidth}}
 \toprule
\textbf{Metrics} & \textbf{LoRaWAN} & \textbf{SigFox} & \textbf{\lte} & \textbf{Wi-Fi}   \\            

\midrule

Max-payload length & 1 - 222 bytes  & 1-12 bytes & UDP/TCP/IP& UDP/TCP/IP \\                                   \midrule
Sending continuous data & 0.165 ms  & 10.5 s & 1-100 ms & 1-100 ms  \\ 
\midrule
Latency & 24 - 2800 ms + 1-100 secs  & 1 - 4.5 s & 500 ms (avg)  & 8 ms (avg)\\ 
\midrule
Throughput  & 250 - 11000 bps   & UL: 100 bps\newline  DL: 600 bps & NB-IoT: UL: 66 kbps; DL: 26 kbps \newline LTE-M: DL: 300 kbps; UL: 380 kbps& Local: 3550 Kbps\newline Remote: 770 Kbps\\ \midrule
Time to connect to network & OTAA: 5.6 s\newline ABP (join not required) & 1-100 ms             & With LTE Reset: 20 s\newline Without LTE Reset: 15.5 s                          & 7.7 s                                             \\  \bottomrule
\end{tabular}

}
\caption{Baseline metrics summary.}
\label{tab:baseline_summary2}
\end{table*}

\subsubsection*{\textbf{Maximum Payload Length}}
Maximum payload size determines how much information (in bytes) can be sent in one message and helps to determine the suitability for an IoT application. For \textit{\lora}, the max payload size varies from 51 bytes to 222\footnote{The payload size is 222 bytes when the device is a repeater and requires optional FOpt control field~\cite{LoRaAlliance2017}.}/242 bytes based on the configuration settings. On the other hand, \textit{\sigfox} allows an \ac{UL} payload of up to 12 bytes and a limit of up to 140 messages per day bytes payload with a limit of 4 messages per day \ac{DL}. Most suitable from the payload perspective, is \textit{\nbiot/\wifi}. LTE \ac{TBS} can support a maximum of 85 bytes DL and 125 bytes UL. However, as TCP/UDP protocol is used in \wifi/\nbiot, the payload is sent as multiple packets (the size and number of which depend on the path \ac{MTU}). So, the payload length for \nbiot/\wifi is bounded by the memory assignment capability of the device.

\begin{table}
\centering
\makebox[0.48 \textwidth][c]{       
\resizebox{0.48 \textwidth}{!}{   

\begin{tabular}{|m{2cm}|m{2cm}|m{2cm}|m{2cm}|m{2cm}|}
\hline
{\small{\textbf{Configuration}}} &
{\small{\textbf{Bitrate (bits/sec)}}} & 
{\small{\textbf{Max payload size (bytes)}}} & 
{\small{\textbf{Time on Air (ms)}}} & 
{\small{\textbf{Max number of messages/day}}}
\\ \hline
{SF12/125} & {250} & {51} & 2793.5 & 12\\ \hline
{SF11/125 kHz} & {440} & {51} & 1560.6 & 23\\ \hline
{SF10/125 kHz} & {980} & {51} & 698.4 & 51\\ \hline
{SF9/125 kHz}  & {1760} & {115} & 676.9 & 53\\ \hline
{SF8/125 kHz}  & {3125} & {222} & 655.9 & 54\\ \hline
{SF7/125 kHz}  & {5470} & {222} & 368.9 & 97\\ \hline
{SF7/250 kHz}  & {11000}& {222} & 184.4 & 195\\ \hline
\end{tabular}
} 
} 
\caption{LoraWAN airtime for max payload in Europe~\cite{LoRaWANAirtime}.}
\label{table:iot_lora}
\end{table}

\autoref{table:iot_lora} represents the max payload sizes with max number of messages per day at different \ac{SF}/bandwidth and respective airtime for \lora~\cite{LoraAirtime}. We use \ac{TTN}, a public community network having a fair access policy~\cite{LoraDutyCycle} that limits the \ac{UL} airtime to 30 seconds per day per node and the \ac{DL} messages to 10 messages per day per node. The max number of messages in \autoref{table:iot_lora} is calculated based on the 1 percent duty and the fair usage policy with maximum payload message. Further, to utilize application payloads efficiently, \lora best practices~\cite{LoraBestPractices} to limit application payloads can be referred.
\sigfox provides \ac{LQI}~\cite{SigfoxLQI} based on the \ac{RSSI} and number of base stations that received a message. However, as only four \ac{DL} messages per day are allowed, it is advisable to set up an HTTP/Email callback to get service-related information.

\subsubsection*{\textbf{Inter-message gap}}
We conducted this primitive experiment to understand the limitation of the time between sending two consecutive messages. For \textit{\lora}, on average it took $0.165~ms$ to send a message. For \textit{\sigfox}, in terms of sending a continuous message on Pycom \fipy end-device, it takes around an average of $10.5~s$, with the minimum $9~s$ and maximum $12~s$ to send a message on \sigfox in RC1 region with $100~bps$. Suppose the application priority is to send the messages fast. In that case, sending a message via \wifi and \nbiot takes a few milliseconds.

To experiment, for \lora, we sent 40 messages with different payloads, ten messages with four SF options offered by \lora, i.e., (SF7 - 1 byte, SF12 - 1 byte, SF7 - 242 bytes, SF12 - 51 byte). For \sigfox, a message with a 12-bytes payload takes $2.08~s$ over the air with a rate of $100~bps$. Further, the device emits a message on a random frequency and then sends two replicas on different frequencies and time~\cite{Sigfox2017a}. We experimented with sending continuous data on \sigfox of variable length starting from $1$ byte to $12$ bytes. We measured the time before sending the message using \texttt{`socket.send(msg)'} and after that. We sent $60$ ($5 \times 12$) messages, three times on average LQI, and one time each on good and excellent LQI.

\subsubsection*{\textbf{Latency}}
We define latency as the delay between transmitting a packet and its arrival at its destination. It combines transmission, propagation, and processing time at both ends.
For \textit{\lora}, \ac{TTN} latency ranges between $24~ms$ (smallest payload - fastest bit-rate) to $2800~ms$ (max-payload on slowest bit-rate) from the end-device to the gateway.
\begin{table}[ht]
\begin{adjustbox}{width=\columnwidth}
\makebox[0.5 \textwidth][c]{       
\resizebox{0.5 \textwidth}{!}{   

\centering

\begin{tabular}{|m{2cm}|m{2cm}|m{2cm}|m{2cm}|}
\hline
 & \textbf{Stated} & \textbf{Observed} & \textbf{Observed} \\
\hline
\textbf{Payload Length} & \textbf{Approximate (sec)} & \textbf{Average (sec)} & \textbf{Good/Excellent (sec)}\\ \hline
\textless{}1 bit        & 1.1                        & 2   &  1         \\ \hline
2 bit - 1 byte          & 1.2                        & 1.6 & 2.0        \\ \hline
2-4 byte                & 1.45                       & 2.3 & 2.1        \\ \hline
5-8 byte                & 1.75                       & 4.5 & 2.5          \\ \hline
9-12 byte               & 2                          & 4.5 & 3            \\ \hline
\end{tabular}
} 
} 

\end{adjustbox}
\caption{\sigfox Payload time approximate time provided by \sigfox~\cite{SigfoxPayload} and observed for Average, Good/Excellent Quality at RC1 Region.}
\label{tab:sigfox-payloadtime-observed}
\end{table}
For \sigfox, \autoref{tab:sigfox-payloadtime-observed} provides the approximate time taken by the message to reach \sigfox backend from the edge device provided by \sigfox~\cite{SigfoxPayload} and observed time taken by payload of different sizes at different LQI (average/good/excellent) in RC1 region for \sigfox. For \nbiot, on average, it has a latency of $576~ms$. It is important to mention that when ping is used the first time, the latency is high in the range of $10~s$ and then stabilises slowly (after $5-10$ pings) to the range of $500-800~ms$. From literature, \nbiot latency ranges around $1-10~s$~\cite{difflte} depending on normal coverage or extended coverage. Latency in LTE-M is around $100-150~ms$ in normal coverage. 
\begin{figure}	
\begin{tikzpicture}
	\begin{axis}[	
		legend style={font=\tiny,at={(1,1)},anchor=north east,column sep=1ex},	
		xlabel = {Ping request number},	
		ylabel = {Time (ms)},
		label style={font=\small},
		xmin=0,ymin=-100,ymax=1600,xmax=100,
    ]	
    \pgfplotsset{every tick label/.append style={font=\small}}
    \pgfplotsset{every axis plot/.append style={thick}}
    \pgfplotsset{%
    width=.45\textwidth,
    height=.3\textwidth
}
	\addplot table[x=Sequence,y=Time,col sep=comma,mark=none]{plot/Latency/wifi_pinglocal.csv};
	\addplot table[x=Sequence,y=Time,col sep=comma,mark=none]{plot/Latency/wifi_pingremote.csv};
	\addplot+[gray, mark=*,mark options={fill=none}] table[x=Sequence,y=Time,col sep=comma ]{plot/Latency/lte_ping_gateway2.csv};
    \legend{Wi-Fi Local Machine, Wi-Fi Remote Machine, NB-IoT Gateway}	
	\end{axis}	
\end{tikzpicture}
\caption{Latency results for pinging a local machine and cloud machine via Wi-Fi and gateway via NB-IoT network.}
\label{fig:wifi_latency}
\end{figure}
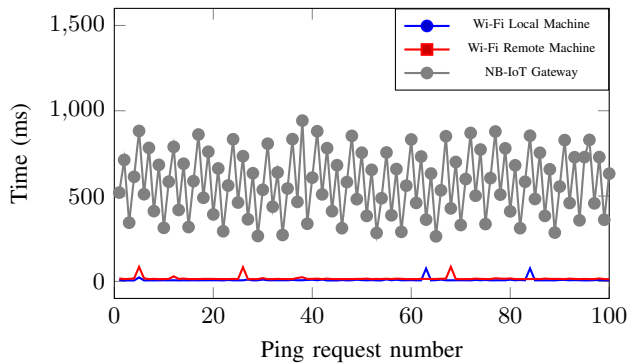
For \textit{\wifi}, latency has an average of $8.32~ms$ and $16.70~ms$ with a standard deviation of $9.93~ms$ and $12.19~ms$ for the machine in local and remote networks, respectively. \autoref{fig:wifi_latency} provides latency of the \wifi network when \fipy pings a machine in the same local network and remotely in the cloud network. The network latency of \nbiot varies significantly compared to the \wifi.

For \textit{\lora}, the transit time from the gateway to the application completely depends on the solution implemented. On \ac{TTN} and with a gateway connected through wired Ethernet, it will take tens of milliseconds (at the current load levels). If the gateway uses a slow cellular connection, the delay will increase. Further, up to a few seconds can add up based on the selected callback mechanism (HTTP, AWS IoT, others). At a high level, latency would be a sum of time-on-air, gateway to network server network latency, duplication window, routing services processing time, a selected callback to application network latency. \lora \ac{TTN} fair usage policy only allows at most $10$ \ac{DL} messages. If we also consider the \ac{DL} latency, one or two seconds could be added to the latency as there are two receive windows after a \ac{UL} message. For \textit{\sigfox}, to understand the latency, we calculated the time when we started sending the message using the device and when it was received at the \sigfox backend. We synced the end-device time using \ac{NTP} with \verb!pool.ntp.org! server. For 
\textit{\nbiot}, we connected to the \nbiot Vodafone network with Pycom provided \ac{SIM}~\cite{VodafoneSIM} having pycom.io \ac{APN}. We figured out our IP Address using AT command `AT+CGCONTRDP' and sent around $100$ ping requests to the gateway, which was three hops away (calculated from TTL). For \wifi, we connected the end-device \fipy to the home \wifi network and calculated the latency by sending 100 $uping$~\cite{uping_gist, uping_usage} requests to a local machine in the same network and to a remote machine on a cloud.

\subsubsection*{\textbf{Throughput}}
This experiment measured the average throughput (bits per second) achieved on each network individually. For \textit{\lora}, bit-rate depends on the bandwidth and SF. In Europe, the regional parameters~\cite{lora_frequency} allow a bandwidth of $125~KHz$ to $250~KHz$ and SF of $7-12$~\cite{LoRaAlliance2017}. \lora data rates range from $0.3~Kbps$ to $50~Kbps$~\cite{LoRaAllianceTechnicalCommitee2017}. 
\begin{table}[h]
\begin{adjustbox}{width=\columnwidth}
\centering
\begin{tabular}{|r|c|c|}
\hline
{Frequency (MHz)} & { \textbf{RC1/RC3/RC5}} & { \textbf{RC2/RC4}} \\ \hline
 
{Uplink center} & {868.130/923.200/923.300} & {902.200/920.800} \\ \hline
{Downlink center} & { 869.525/922.200/922.300} & { 905.200/922.300}  \\ \hline
{Uplink data rate (bit/s)} & {100} & { 600} \\ \hline
{Downlink data rate (bit/s)} & {600} & { 600} \\ \hline
\end{tabular}
\end{adjustbox}
\caption{Sigfox radio configuration~\cite{SigfoxRC}.}
\label{tab:sigfox-rc}
\end{table} 
For \textit{\sigfox}, \autoref{tab:sigfox-rc} provides \ac{UL} and \ac{DL} frequency and data rate for different regions. Based on the \sigfox frequency, the data rate could be determined. For \textit{\nbiot}, data rate~\cite{difflte} is $26~Kbps$ in the \ac{DL}, and $66~Kbps$ in the \ac{UL}. LTE-M has approximately $300~Kbps$ in \ac{DL} and $380~Kbps$ in the \ac{UL} in half-duplex. On an average on the field, $100$ to $150~Kbps$ are reached in both directions.
\begin{figure}	
\begin{tikzpicture}	
	\begin{axis}[	
		legend style={font=\tiny,at={(1,0.6)},anchor=north east,column sep=1ex},
		xlabel = {Time interval (secs)},	
		ylabel = {Kbits/sec},
		cycle list name=black white,
        every axis plot/.append style={ultra thick}
    ]
    \pgfplotsset{every tick label/.append style={font=\small}}
    \pgfplotsset{every axis plot/.append style={thick}}
    \pgfplotsset{%
    width=.45\textwidth,
    height=.25\textwidth
}
	\addplot [only marks,mark=x,mark options={scale=1}] table[x=Sequence,y=Bandwidth,col sep=comma,]{plot/Bandwidth/dtlocal.csv};
	\addplot [mark=x,mark options={scale=0.5}] table[x=Sequence,y=Bandwidth,col sep=comma]{plot/Bandwidth/dtremote.csv};
    \legend{Local Machine, Remote Machine}	
	\end{axis}	
\end{tikzpicture}
\caption{Bandwidth results using iperf when running on local network and cloud.}
\label{fig:wifi_bandwidth}
\end{figure}
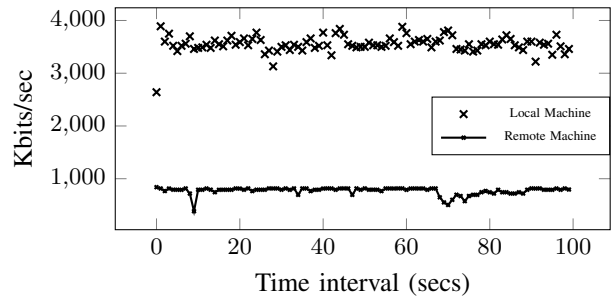
For \wifi, bandwidth has an average of $3550.8~Kbps$ and $770.181~Kbps$ with a standard deviation of $157.19~Kbps$ and $71.86~Kbps$ when $iperf3$ is hosted locally in the local network and cloud network, respectively. \autoref{fig:wifi_bandwidth} provides the bandwidth of the \wifi network when \fipy pings and connects to the \texttt{iperf} server in the same local network and remotely in the cloud network. 

For our experiments, for \nbiot, we are using Pycom provided Vodafone SIM; the User Equipment (UE) can only communicate to a white-listed IP address because of which we were unable to host an instance of \texttt{iperf} on a server and calculate throughput. For \textit{\wifi}, Pycom \fipy utilises \texttt{ESP32} which provides $20~Mbps$ TCP RX/TX in the test~\cite{esp32_wifi} performed in the lab. The bandwidth and throughput was calculated using \texttt{uiperf3}~\cite{uiperf3}.

\subsubsection*{\textbf{Time to connect to the network}}
\label{subsubsec_time_to_connect}
 We conducted baseline experiments to understand the connection time an end device takes to join the different networks. It helps to estimate switching overhead if the IoT devices need to switch from one network technology to another. \textit{\lora} allows activation by two methods Over-the-Air Activation (OTAA) and Activation by Personalisation (ABP). OTAA took on an average $5.6~s$ with a minimum $4~s$ to a maximum $7~s$ to join the network. On the other hand, ABP provides hard-coded session keys and allows the sending of data without joining. In case of an emergency where the device tries to send only one message and is unsure about the coverage of \lora, the message can be sent using maximum SF12 to have a maximum range. For \textit{\sigfox}, creating a socket for \sigfox taken an average of few $ms$. 
 For \nbiot, we present  the timings for the different methods in \autoref{tab:nb-iot-connect}.
When the LTE modem is connected to the network first time, it takes a significant amount of time to connect to the network as it searches, registers itself to the network, it could take approximately $15~mins$ to $60~mins$ to attach to the network, whereas \wifi takes approx 5.6 seconds to connect to the specified network.
\begin{table}[ht]
\begin{adjustbox}{width=\columnwidth}
\begin{tabular}{|l|l|l|l|l|l|l|l|l|}
\hline
\textbf{NB-IoT} & \textbf{} & \textbf{Reset} & \textbf{Init} & \textbf{Attach} & \textbf{Connect} & \textbf{Disconnect} & \textbf{Deattach} & \textbf{Deinit} \\ \hline
With Reset      & Avg       & 6.37           & 0             & 12.64           & 1.29             & 7.23                & 1.13              & 0.09            \\ \hline
                & Min       & 6              & 0             & 11              & 1                & 7                   & 0                 & 0               \\ \hline
                & Max       & 7              & 0             & 17              & 2                & 8                   & 2                 & 1               \\ \hline
Without Reset   & Avg       & -              & 2.07          & 12.15           & 1.31             & 7.23                & 0.98              & 0.15            \\ \hline
                & Min       & -              & 1             & 8               & 1                & 7                   & 0                 & 0               \\ \hline
                & Max       & -              & 3             & 20              & 2                & 8                   & 2                 & 1               \\ \hline
\end{tabular}
\end{adjustbox}
\caption{NB-IoT connect times in seconds.}
\label{tab:nb-iot-connect}
\end{table}

We conducted a baseline experiment for Lora to understand the connection time an end device takes to join the \lora network through OTAA and repeated it 24 times. For \nbiot, we conducted experiments to measure the time taken for initialisation, attach, connect, detach, disconnect, deinit and modem reset. We performed two experiments - one when LTE modem is reset before initialisation and one without the reset. LTE allows \ac{PSM} by configuring the period how often the device will connect and how long it will stay actively connected. During the sleep, the LTE modem will go into a low power state, but it will stay attached to the network; thus, no time is spent for attaching after waking up. For \textit{\wifi}, to understand the time taken to connect to \wifi, we calculated the time taken for \wifi init, scan, connect, disconnect, deinit. We experimented 80 times and found that it takes approximately 2.1 seconds to scan the \wifi networks and approximately 5.6 seconds to connect to the specified network. \wifi init, disconnect, and deinit were almost instantaneously in the range of milliseconds.

\subsubsection*{\textbf{Time to reconnect to \wifi, Internet}}
To understand how much time an IoT device takes to reconnect with the \wifi and the internet. We connected a \ac{SCK}~\cite{SCK}, Pycom FiPy to Home \wifi, a machine via Ethernet to the home router and turn-off-on the \wifi. We created a python script that pings the three hosts: the router, the IoT device (SCK, FiPy), and the cloud machine (google.com) and provided time between the device going offline and coming online. It took approx 1 min 16 sec, 1 min 40 sec, 3 min 5 seconds to get the connectivity back to the router, IoT device, and the internet.




\section{Evaluation and Discussions}
\label{sec:evaluation}
In \autoref{sec:resourcemanagement}, we have shown that both variants of the proposed resource management algorithm ($CABF$ and $CABF_{inv}$) perform better than the baseline bin-packing algorithms we considered. In this section, we implemented one of the variants, namely $CABF_{inv}$, as part of a multi-network resource allocator running over the \fipy platform. We then performed a number of experiments to evaluate the algorithm performance over an edge-node prototype following the experiment setup as shown in the Figures~\ref{fig:exp_block} and \ref{fig:exp_setup}. The choice of the $CABF_{inv}$ was made in order to try to provide service to all message flows (rather than focus on increasing service for the most critical flows, which would perhaps be the goal in a real deployment) for the sake of demonstrability, i.e. so we can show the sharing of the network interfaces by several flows operating at different levels of criticality. 

\subsection{Criticality-aware allocation of network resources using $CABF_{inv}$}
In this section, we show the performance of the proposed $CABF_{inv}$ algorithm when allocating network resources to application flows in a criticality-aware manner. We consider the same application flows and QoS requirements presented in Section \autoref{sec:system-model} and follow the approach described in Section \autoref{sec:resourcemanagement} where application flows can request service at different criticality levels from the multi-network resource allocator running at \fipy. The multi-network resource allocator running $CABF_{inv}$ algorithm provides service according to those requirements while considering the network availability conditions, so in this experiment we consider a number of realistic scenarios and evaluate the percentage of served requests and the corresponding criticality levels they were assigned.

In this experiment we consider the four available networks have following maximum bandwidths  \wifi($750~Kbps$), \nbiot \ac{UL} ($55~Kbps$),  LoRa SF7-125KHz ($5.47~Kbps$) and \sigfox \ac{UL} ($100~bps$).
\begin{table}[]
\makebox[0.5 \textwidth][c]{       
\resizebox{0.48 \textwidth}{!}{   
\begin{tabular}{c|c|c|c|c|c|c|c|c|c|c|}
\cline{2-11}
\textbf{}                                                  & \multicolumn{8}{c|}{\textbf{Message Flows}}               & \multirow{1}{*}{\textbf{\begin{tabular}[c]{@{}c@{}}\% flows\end{tabular}}} & \multirow{1}{*}{\textbf{\begin{tabular}[c]{@{}c@{}}avg\end{tabular}}} \\ \cline{2-9}
    & 1      & 2      & 3    & 4    & 5    & 6    & 7    & 8    &  \textbf{served}   & \textbf{crit}                                                 \\ \cline{1-9}
\multicolumn{1}{|c|}{\textbf{\begin{tabular}[c]{@{}l@{}}Requested \\ Criticality level\end{tabular}}} & 1,2,3  & 1,2,3  & 1,2  & 1,2  & 1,2  & 1,2  & 1,2  & 1    &    & \textbf{level}                   \\ \cline{1-9}
                                        
\multicolumn{1}{|c|}{\textbf{\begin{tabular}[c]{@{}l@{}}Network \\ Interfaces\end{tabular}}}          & \multicolumn{8}{c|}{{\textbf{Allocated Criticality Level}}}                                     &                                          &                                                 \\ \hline
\multicolumn{1}{|c|}{Wi-Fi}                                & 1      & 1      & 1    & 1    & 1    & 1    & 1    & 1    & 100                                      & 1                                               \\ \hline
\multicolumn{1}{|c|}{LoRa}                                 & 1      & 1      & 1    & 2    & 1    & 2    & 1    & 1    & 100                                      & 1.25                                            \\ \hline
\multicolumn{1}{|c|}{NB-IoT}                               & 1      & 1      & 1    & 1    & 1    & 1    & 1    & 1    & 100                                      & 1                                               \\ \hline
\multicolumn{1}{|c|}{Sigfox}                               & 2      & 2      & 1    & 2    & 1    & 2    & 2    & 1    & 100                                      & 1.625                                           \\ \hline
\multicolumn{1}{|c|}{Wi-Fi + LoRa}                         & 1\#    & 1\#    & 1\#  & 1*   & 1\#  & 1*   & 1\#  & 1\#  & 100                                      & 1                                               \\ \hline
\multicolumn{1}{|c|}{Wi-Fi + Sigfox}                       & 1\#    & 1\#    & 1+   & 1\#  & 1+   & 1\#  & 1\#  & 1+   & 100                                      & 1                                               \\ \hline
\multicolumn{1}{|c|}{NB-IoT + LoRa}                        & 1*     & 1*     & 1*   & 1-   & 1*   & 1-   & 1*   & 1*   & 100                                      & 1                                               \\ \hline
\multicolumn{1}{|c|}{NB-IoT + Sigfox}                      & 1-     & 1-     & 1+   & 1-   & 1+   & 1-   & 1-   & 1+   & 100                                      & 1                                               \\ \hline
\end{tabular}

} 
} 
\caption{Obtained criticality level (1 $|$ 2 $|$ 3) and network allocation (* Wi-Fi $|$ \# LoRa $|$ + Sigfox $|$ - NB-IoT) for motivating example in FiPy.}
\label{tab:AssistedLivingExampleResultsFiPY}
\end{table}

\autoref{tab:AssistedLivingExampleResultsFiPY} shows the allocated criticality level, percentage of flows served and average criticality level for the messages flows defined in \autoref{tab:AssistedLivingExample}. Each row of the table shows the metrics obtained by running the $CABF_{inv}$ algorithm over a different network scenario. Scenarios include situations such as when only a single network is available (only \wifi, \lora, \nbiot or \sigfox) or when two different networks are available (such as \wifi or \nbiot with \lora and \sigfox). When a high-bandwidth network such as \wifi and \nbiot is available, we can see that $CABF_{inv}$ is able to assign the lowest criticality level to all flows and to provide all of them with service. We also observe that when only low-bandwidth network interfaces are available (e.g. \sigfox), all flows are still serviced but the average allocated criticality is higher (i.e. flows are only allowed to use the network under more constrained levels of service). Average criticality level is calculated as the sum of all the assigned criticality level divided by the number of message flow allocated.

\subsection{Time complexity and Context Switching of $CABF_{inv}$ algorithm}
\label{subsec:time_complexity}
We measured the running time of $CABF_{inv}$ algorithm on the \fipy board and the \rpi,  repeated it ten times, and average run on \fipy takes $1300 ms$ whereas on \rpi it takes $7.1 ms$.

When a networking event (such as \wifi is disconnected), the allocation algorithm $CABF_{inv}$ has to be executed again. This results in time delay due to de-allocation of old message flows and allocation of new message flows. During this time delay, there's a possibility that the \rpi would have written a message on the UART. 

As there is a possibility that by the time, Message Flow Element Allocation (MFEA) message was received by \rpi, the previous running threads (simulating message flows) would have written few messages to the UART. To resolve this, before doing the re-allocation, we send a message \texttt{<INFO:RE-ALLOC:INIT>} to \rpi that, we are going to do the re-allocation, stop sending any message to the UART to minimise the loss of messages. On receiving that message, \rpi pauses all the current threads of message flow. Further, \fipy store the old allocations and until it receives an acknowledgement message \texttt{<INFO:RE-ALLOC:ACCEPTED>} from the \rpi that it has received the MFEA, it keeps allocating using previous allocation (except the network interface which was lost).

In this case, we log the time, when \texttt{RE-ALLOC:INIT} message was written to the UART by \fipy initiating re-allocation, the time \rpi received MFEA message flow allocation message from \fipy, and the time taken by \rpi to stop all previous threads (which are simulating the message flows) and generate new threads (as per new allocation). We calculated the time for context switching as the time difference between re-allocation \texttt{init} message written by \fipy and the re-allocation accept message received by \fipy.  This whole context switching takes $1.3~s$ to $1.5~s$ which includes stopping thread, creating new threads, re-alloc \texttt{init} message, re-alloc MFEA time from \rpi to \fipy and re-alloc accept from \fipy to \rpi.

\subsection{Discussions}
Our work also has certain limitations. Firstly, the CABF algorithm currently does not handle network dynamics such as a change in network bandwidth due to dynamic change of wireless channel and link conditions. The preliminary decision about the network capacity is based on the network availability (whether the network is available or not), and the algorithm calculates the network bandwidth at the start of the network connection.
Currently, it is difficult to generate or simulate network problems during application communication to evaluate the consequences on the flows (latency, loss, throughput). For instance, currently, \fipy does not provide the \wifi callback function~\cite{FiPy_Callback} and does not provide any way to know that \wifi is disconnected. In LTE, we can remove the SIM card or the LTE antenna during a stable connection to simulate network connectivity loss. However, removing SIM or antenna is not officially recommended as they can cause damage to the device. Regarding generating network loss in Lora and Sigfox, both are stateless. \fipy provides a way to check if the device has joined \lorawan; however, no way to find whether it is still connected or not.
Because of the above reasons,  to simulate the loss of \wifi, we have manually set the \wifi bandwidth to zero and then called the re-allocation function. The multi-network resource allocator successfully allocates the message flows to the available network interfaces. From the network bandwidth perspective (change in bandwidth due to network conditions), a for loop that checks for the LoRa \ac{SF}, \wifi, and \nbiot bandwidth at regular intervals can be implemented. However, it requires better support for threading. We will eventually implement the features based on the device support for \wifi callback in the future.

Secondly, there are few device limitations. FiPy does not provide \wifi callback to indicate if the device got disconnected from the \wifi network. Currently, when FiPy is connected to both \wifi and NB-IoT simultaneously, it does not provide a way to define the network interface to be used for sending the packet. Further FiPy team does not advise using both networks simultaneously to simulate a WiFi-LTE bridge, as it will be very slow and expensive~\cite{FiPy_WiFiNBIoT}.

Thirdly, currently, we take a set of message flows and allocate them all together. Because of this, old message flows are de-allocated and re-assigned with either the same or different criticality levels. In future work, we will provide the capability to allow an application to define a new message flow and allocate it from the existing networks without de-allocating and re-allocating the old ones.

Further, there are different industrial products~\cite{multiIndus, multiIndus2} in the market that provide communication via multiple radio interfaces (such as \wifi, 4G, \lora, \lte). However, either they provide only \lora or \lte with \wifi. Currently, we are only aware of \fipy that provides multi-network connectivity for \lora, \lte, \sigfox, \wifi, and Bluetooth. Further, our work enables criticality-aware applications to send messages by allocating resources (network) per the criticality level and network availability. The transmission range of \wifi and other WPAN is different, and it is possible to assign the communication resources to different types of traffic. There can be different factors for consideration in the case of multiple radio devices, e.g., bandwidth, delay, rate adaptation, IP support, and others. Currently, our work considers bandwidth and availability to ensure that applications can send messages as per the defined criticality level.

With the development and popularization of 4G/5G networks, the IoT edge has also shown more possibilities in IoT, VR, and AI intelligence. In this context, NB-IoT, LoRa, and Sigfox provide low-bandwidth network communication methods that are very limited. There might be a case where LPWAN might seem insignificant. On the other hand, our work targets critical edge applications that need to work even when high-bandwidth networks are unavailable.

\section{Related Work}
\label{sec:related_work}
This section presents related work that crosses the intersection of  LPWAN, edge resilience, and  ILP (Integer Linear Programming) formulations for IoT and edge computing.

For example, Chaudhari \etal~\cite{Chaudhari2020} provided a comprehensive survey on various LPWAN technologies and presented these technologies concerning application requirements, such as coverage, capacity, cost, low power, and deployment complexity, and provided a comprehensive survey on both standard and non-standard LPWAN technologies.

From the use case perspective, Santos \etal~\cite{Santos2018} evaluated LPWAN technologies for air quality application during ``City of Things'' project that can be used as wireless communication enablers for the smart city use case. Further, it also performed anomaly detection for smart city applications using different unsupervised and outlier detection algorithms.  Roque  \etal~\cite{Roque2020} created a prototype to detect fire detection in outdoor environments (forests) based on LPWAN networks (\sigfox) and temperature and gas sensor measurements. Rubio-Aparicio \etal~\cite{Rubio-Aparicio2019} implemented an LPWAN residential water management solution supported by hybrid IoT LoRa-Sigfox architecture. All above solutions use provides some level of resiliency by sending data on Lora and Sigfox, without guaranteeing applications' QoS requirements.


ILP formulations have been used for resource provisioning for IoT applications at the gateway level; for example, Santos \etal~\cite{Santos2021} presents a MILP (Mixed ILP) formulation for resource provisioning in Fog computing, taking into account the \ac{SFC} concepts, different LPWAN technologies (LoRa, IEEE 802.11 ah), and multiple optimization objectives. It considers end-to-end system into three segments  - sensors/things level, gateways/routers (Fog), and the cloud and presents smart-city use-cases for garbage collection, air quality monitoring, and \ac{CCTV} monitoring. Kim \etal~\cite{Kim2020} use ILP formulation to create secure migration policies for the communication between things (sensors) and a trusted edge system providing authentication services in the event of \ac{DoS} attacks or failures, resulting in resilient authentication and authorization for IoT. Our work shows that IPL can also be used at the IoT device level to optimize the latency and resiliency of different applications using a Multi-communication network.

From the perspective of improving resilience, Qin \etal implemented~\ac{MINA}~\cite{Qin2013,Qin2014,Qin2014sdn} a reflective \ac{OAA} middleware approach to manage dynamic and heterogeneous multi-network (such as \zigbee, \bluetooth, PANs, MANETs, 3G/4G, WLAN) in pervasive environments to ensure reliable communication for end applications. They presented a formal analysis that can guide network administrators in their decisions to adapt network configurations to achieve mission or application objectives proactively. Compared to this work in our paper,  we analyzed seamless switching of the networks on a hardware testbed to meet the resiliency requirements. We provided seamless switching while maintaining the critical application requirements.

The SCALE2~\cite{Uddin2016} leveraged MINA and implemented a multi-tier and multi-network approach to drive data flow from IoT devices to cloud platforms. The authors implemented a local \ac{SDN}-enabled the network, which is adaptive to the network changes to which IoT client devices are connected. This solution's architecture and deployment examples used separate adapters (device)  for each communication radio, thus needing another computing device to run the SCALE client software. However, in our work, we use all radios integrated on a single board to allow fast switching between networks on the device level.


Wider aspects of resilience has been discussed in mission-critical applications like autonomous driving, tactile healthcare, and public safety. For example, Modarresi \etal~\cite{Modarresi2020} presented a graph-theoretical approach to model IoT systems in smart homes with integrated heterogeneous networks and explored resilience properties. Similarly, Chaterji \etal~\cite{Chaterji2019} presents the resilience of \ac{CPS} and discusses two techniques resilience-by-design and resilience-by-reaction. Harchol \etal~\cite{Harchol2020} proposed a framework to improve edge-computing resilience for session-oriented applications. They utilized message replay and checkpoint-based mechanisms to make client-edge-server systems more tolerant to edge failures, client mobility. Carvalho \etal~\cite{Carvalho2020} implement a replication mechanism LoRa-REP for replicating critical messages on LoraWAN by sending them at different SF and improving redundancy in LoRaWAN for mixed-criticality scenarios.

From the literature, this is evident that different forms of replication and redundancy mechanisms are used to achieve resilience in the networks, however, none of those as mentioned above work used different LPWAN and \wifi as seamless multi-network infrastructure at the device and the Edge network to meet the guaranteed message delivery.

Our work focuses on achieving network resiliency using the LPWAN network on resource-constrained end devices by providing the capability to the end device to evaluate the application requirements and select the suitable network medium while allowing graceful degradation of services in the event of failures. Further, our work implements an ILP solver in micro-python that can run on a resource-constrained device. Also, the device having multi-network connectivity has benefits in terms of deployment in mission-critical applications (tactile healthcare, public safety in smart city) and for mobility-based IoT like autonomous driving, for example, if one type of network exists in one area, whereas in another geographical location, there is another network, the application can perform a smooth and seamless network switching.



\section{Conclusion and Future Work}
\label{sec:conclusion}
The resiliency and reliability requirements of IoT applications vary from non-critical (best delivery efforts) to safety-critical with time-bounded guarantees. In this work, we systematically investigated how to meet these applications mixed-criticality QoS requirements in multi-communication networks. 

We presented the network resiliency requirements of IoT applications by defining a theoretical multi-network resource system model and proposed and evaluated a list of resource allocation algorithms and found Criticality-Aware Best Fit ($CABF_{inv}$) algorithm works better to meet high criticality requirements of the example applications. The algorithm provides the best-effort QoS match by taking into consideration the underlying dynamic multi-network environments. 
We analysed and evaluated the bandwidth, latency, throughput, maximum packet size of LPWAN technologies, such as \sigfox, \lora, and \nbiot and implemented and evaluated an adaptive \system system with Criticality-Aware Best Fit (CABF) resource allocation to meet the application resiliency requirements using underlying LPWAN technologies on \rpi and \fipy. 

In the current implementation of \system, we took bandwidth and subsequent inter-message period into consideration for defining criticality\footnote{Upon acceptance, we will publish all our source code and datasets.}. In future, we would like to extend  multi-network resource allocator to include message payload size, message transmission frequency, security, privacy and energy consumption parameters in the allocation algorithm. The new allocator would provide applications more flexibility to choose and optimise their resources and QoS for a multi-communication network.
In summary, we investigated the limits and metrics required for the best-effort high criticality resilience in multi-communication networks. We presented our findings on how to achieve 100\% of the best-effort high criticality level message delivery using multi-communication networks. Our work will help build reliable applications on IoT Edge and provide solutions from the perspective of communication networks to improve service quality and fault tolerance on resource-constrained edge devices. It also opens up new research directions to build reliable and trustworthy IoT applications over robust and resilient IoT Edge.

\section*{Acknowledgements}
Yadav and Kumar are supported in part by "Data Negotiability in
Multi-Mode Communication Networks" project funded by EPSRC
grant EP/R045178/1.

\ifCLASSOPTIONcaptionsoff
  \newpage
\fi



\balance
\bibliographystyle{IEEEtran}
\bibliography{IEEEabrv,hdi.bib}
\end{document}